\documentclass[sigconf, nonacm]{acmart}

\newcommand\vldbdoi{XX.XX/XXX.XX}
\newcommand\vldbpages{XXX-XXX}
\newcommand\vldbvolume{14}
\newcommand\vldbissue{1}
\newcommand\vldbyear{2020}
\newcommand\vldbauthors{\authors}
\newcommand\vldbtitle{\shorttitle} 
\newcommand\vldbavailabilityurl{URL_TO_YOUR_ARTIFACTS}
\newcommand\vldbpagestyle{plain} 

\begin{document}
\title{OVT-MLCS: An Online Visual Tool for MLCS Mining\\ from Long or Big Sequences}


\author{Zhi Wang}
\email{zhiwang@stu.xidian.edu.cn}
\affiliation{
    \institution{Xidian University, Xi'an, China}
}

\author{Yanni Li}
\authornote{Both Yanni~Li and Bing~Liu are co-corresponding authors.}
\email{yannili@mail.xidian.edu.cn}
\affiliation{
    \institution{Xidian University, Xi'an, China}
}

\author{Tihua Duan}
\email{duantihua@163.com}
\affiliation{
    \institution{Yushang  Info. Tech. Co., LTD, China}
}

\author{Bing Liu}
\authornotemark[1]
\email{liub@uic.edu}
\affiliation{
    \institution{University of Illinois Chicago, USA}
}

\author{Liyong Zhang}
\email{zhangliyong@xidian.edu.cn}
\affiliation{
    \institution{Xidian University, Xi'an, China}
}

\author{Hui Li}
\email{hli@xidian.edu.cn}
\affiliation{
    \institution{Xidian University, Xi'an, China}
}

\begin{abstract}
Mining multiple longest common subsequences (\textit{MLCS}) from a set of sequences of three or more over a finite alphabet $\Sigma$ (a classical NP-hard problem) is an important task in a wide variety of application fields. Unfortunately, there is still no exact \textit{MLCS} algorithm/tool that can handle long (length $\ge$ 1,000) or big (length $\ge$ 10,000) sequences, which seriously hinders the development and utilization of massive long or big sequences from various application fields today. To address the challenge, we first propose a novel key point-based \textit{MLCS} algorithm for mining big sequences, called \textit{KP-MLCS}, and then present a new method, which can compactly represent all mined \textit{MLCSs} and quickly reveal common patterns among them. Furthermore, by introducing some new techniques, e.g., real-time graphic visualization and serialization, we have developed a new online visual \textit{MLCS} mining tool, called OVT-MLCS. OVT-MLCS demonstrates that it not only enables effective online mining, storing, and downloading of \textit{MLCSs} in the form of graphs and text from long or big sequences with a scale of 3 to 5000 but also provides user-friendly interactive functions to facilitate inspection and analysis of the mined \textit{MLCS}s. We believe that the functions provided by OVT-MLCS will promote stronger and wider applications of \textit{MLCS}.
\end{abstract}

\maketitle

\pagestyle{\vldbpagestyle}
\begingroup\small\noindent\raggedright\textbf{PVLDB Reference Format:}\\
\vldbauthors. \vldbtitle. PVLDB, \vldbvolume(\vldbissue): \vldbpages, \vldbyear.\\
\href{https://doi.org/\vldbdoi}{doi:\vldbdoi}
\endgroup
\begingroup
\renewcommand\thefootnote{}\footnote{\noindent
This work is licensed under the Creative Commons BY-NC-ND 4.0 International License. Visit \url{https://creativecommons.org/licenses/by-nc-nd/4.0/} to view a copy of this license. For any use beyond those covered by this license, obtain permission by emailing \href{mailto:info@vldb.org}{info@vldb.org}. Copyright is held by the owner/author(s). Publication rights licensed to the VLDB Endowment. \\
\raggedright Proceedings of the VLDB Endowment, Vol. \vldbvolume, No. \vldbissue\ %
ISSN 2150-8097. \\
\href{https://doi.org/\vldbdoi}{doi:\vldbdoi} \\
}\addtocounter{footnote}{-1}\endgroup

\ifdefempty{\vldbavailabilityurl}{}{
\vspace{.3cm}
\begingroup\small\noindent\raggedright\textbf{PVLDB Artifact Availability:}\\
The source code, data, and/or other artifacts have been made available at \url{https://github.com/OVT-MLCS/OVT-MLCS-Tool}.
\endgroup
}

\section{Introduction}
Data in various applications often can be abstracted as character sequences over a finite alphabet $\Sigma$, e.g., DNA and protein sequences in biology. Searching for multiple longest common subsequences (\textit{MLCS}) from a set of given sequences (i.e., those greater than or equal to 3 sequences) over $\Sigma$ is a classical NP-hard problem~\cite{r1}, which is related to the identification of sequence similarity. It has many important applications in bioinformatics, computational genomics, pattern recognition, data mining, information extraction, e.g., for early cancer detection \cite{r2}, cancer diagnosis and treatment \cite{r3}, and COVID-19 virus evolution research \cite{r4}. With the development of the latest fourth--generation sequencing technology \cite{nr1}, the length and size of biological and other types of very long character sequences, called \textit{big sequences} (i.e., their length $\ge$ 10,000), are growing rapidly. Discovering \textit{MLCSs} from such long or big sequences and exploring patterns in a large number of \textit{MLCSs} (hereafter denoted by \textit{MLCSs}) have posed urgent application needs.
\vspace{-0.8em}
\begin{figure}[!h]
    \centering
    \includegraphics[width=0.95\linewidth]{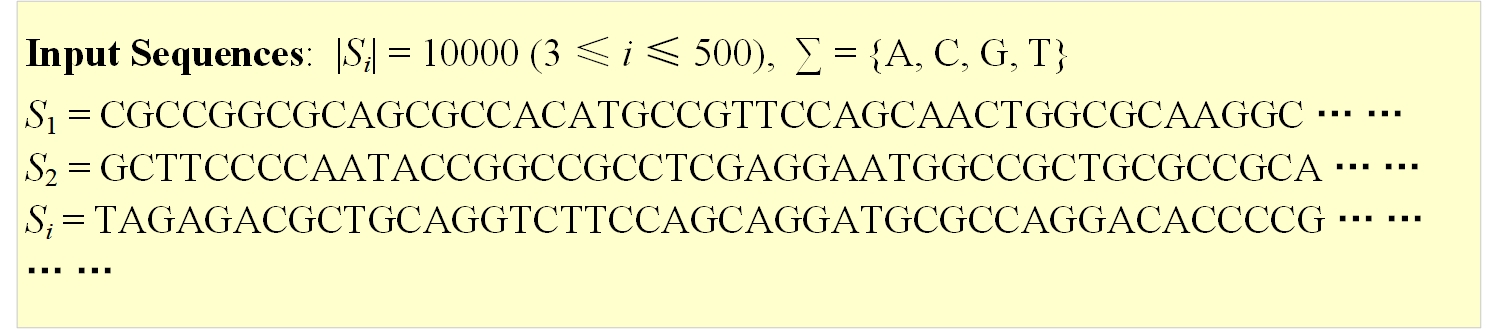}
    \label{fig0}
\end{figure}
\vspace{-1em}

For example, given a set of big DNA sequences $S_1, S_2, S_i, ...$ from the biomedical field shown in the figure above, to perform their similarity comparison for cancer gene common pattern detection and treatment and/or phylogenetic analysis, users may have the following application needs: 1) Effectively and efficiently mine all \textit{MLCSs} of DNA big sequences. 2) online and intuitively observe their \textcolor{black}{\textbf{common patterns}\footnote{Common patterns denote those successive common subsequences in all mined \textit{MLCSs}.}} from a large number of mined \textit{MLCSs}, and obtain accurate statistical information about the \textit{MLCSs}. 3) Effectively and efficiently mine only the top-\textit{k} \textit{MLCSs}, not all massive \textit{MLCSs}, from the input sequences that make sense for their applications. 4) Users can interact in a friendly manner with the mined results to support further observation and analysis.

Unfortunately, although many outstanding \textit{MLCS} algorithms (\textit{e.g.},~\cite{r5,r6,r7}, etc.) and mining tools (\textit{e.g.}, BWA-MEM2{\footnote{\scriptsize https://gitcode.com/gh-mirrors/bw/bwa-mem2}}, BLAST \cite{r10}, 
Clustal Omega \cite{r11}
, etc.) have been proposed/developed in the past forty years, the existing exact \textit{MLCS} algorithms and corresponding tools still face serious challenges as MLCS mining is an NP-hard \textit{MLCS} problem: 1) They cannot handle long or big sequences due to the overwhelming size of their underlying problem-solving graph model \textit{MLCS-DAG}, leading to the issue of memory explosion or extremely high time complexity. 2) They cannot effectively extract patterns from a large number of \textit{MLCSs} as these results, which are outputted one by one, do not have intuitive structural or visual patterns without additional special processing. As a result, \textit{so far, there is no exact \textit{MLCS} algorithm/tool that can mine \textit{MLCSs} from a set of long or big sequences and meet the above users' needs}.

To overcome the challenges, we first propose a novel key point-based \textit{MLCS} algorithm for big sequences mining (see \cite{r8} for details), called \textit{KP-MLCS}, and then present a new method, which can compactly represent all mined \textit{MLCSs} and quickly reveal common patterns of the mined \textit{MLCSs}. By introducing some new techniques, e.g., real-time graphic visualization and serialization, a new online visualization tool \textit{OVT-MLCS} is developed, which can effectively enable online mining, storing, and downloading of \textit{MLCSs} in the form of graphics and text from input sequences (including long or big sequences), and provides user-friendly interactive functions for the user to further inspect and analyze the mined \textit{MLCSs}. These together contribute to meeting the above users' needs. In today's rapid growth of long or big sequences from various fields, we believe that the functions provided by \textit{OVT-MLCS} will promote stronger and wider applications of \textit{MLCS}.

\begin{figure}[!h]
    \centering
    \includegraphics[width=0.90\linewidth]{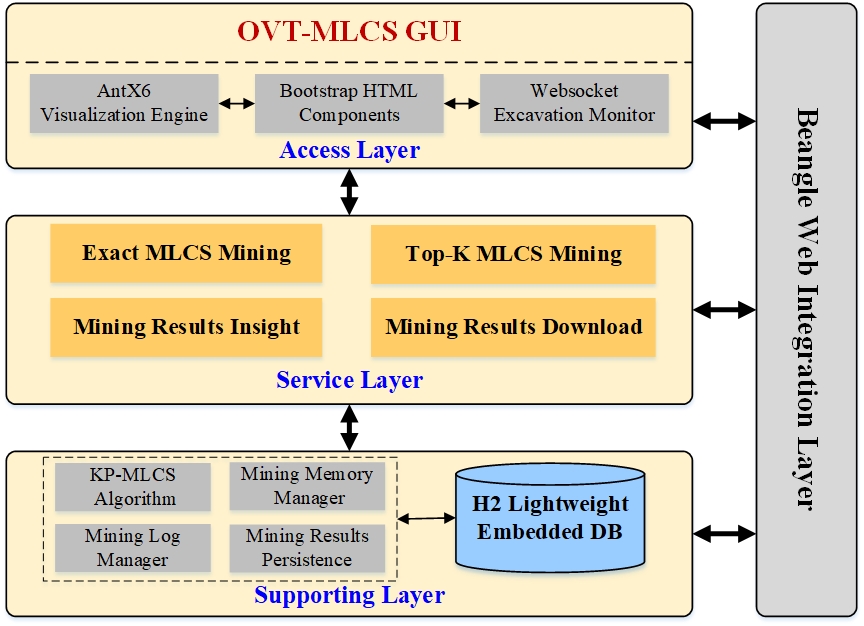}
    \caption{The architecture of OVT-MLCS, where the components with an orange box are functional components provided as user APIs, while the components with a gray box are system-supporting components transparent to users.}
    \label{fig1}
\end{figure}

\vspace{-1.5em}
\section{System Overview}
OVT-MLCS is a lightweight Web application based on open source pure JAVA components AntX6\footnote{\scriptsize https://x6.antv.antgroup.com/}, Bootstrap HTML\footnote{\scriptsize https://getbootstrap.com/}, WebSocket\footnote{\scriptsize https://developer.mozilla.org/en-US/docs/Web/API/WebSocket}, Beangle Web\footnote{\scriptsize https://github.com/beangle/webmvc} and database H2\footnote{\scriptsize https://cloud.tencent.com/developer/article/2333059}. The architecture of OVT-MLCS is depicted in Figure \ref{fig1}, where the functions of the components in the \textit{Supporting Layer} are as their names suggest. The following will focus on discussing the key functional components of OVT-MLCS.

\subsection{OVT-MLCS GUI}
\begin{figure}[!h]
    \centering
    \includegraphics[width=0.95\linewidth]{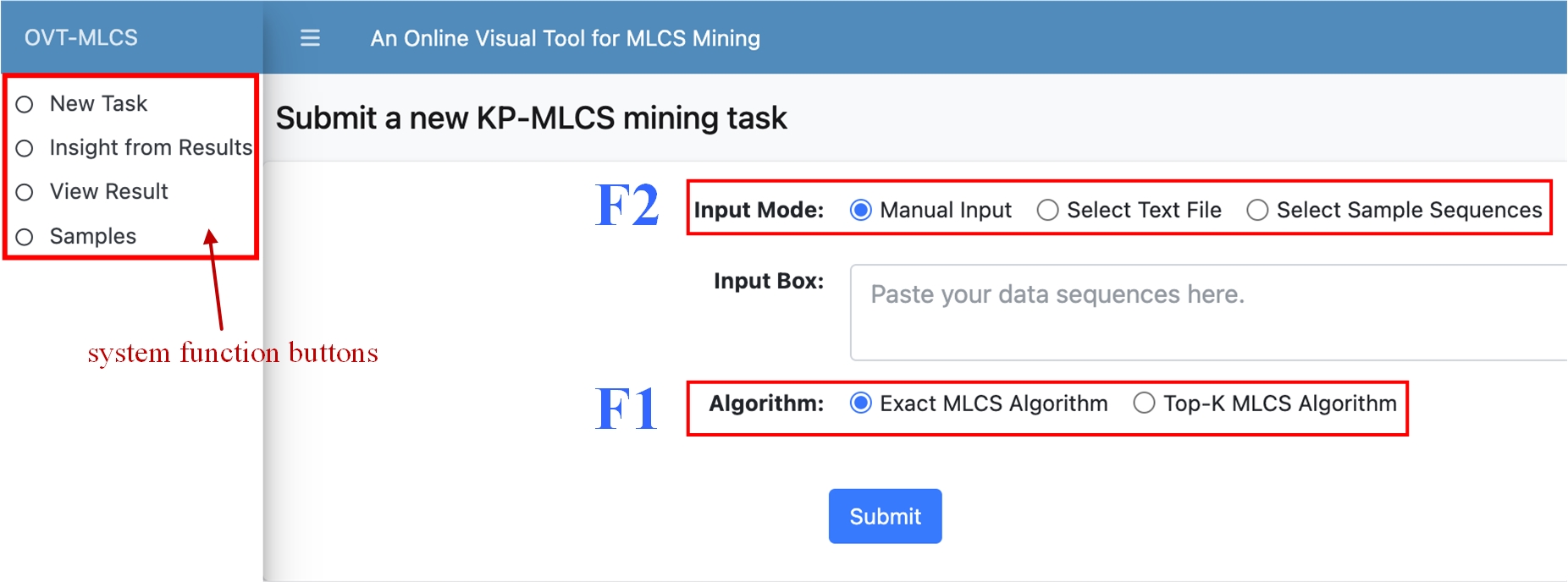}
    \caption{OVT-MLCS graphical/visual user interface (GUI).}
    \label{fig2}
\end{figure}

OVT-MLCS is a browser-based graphical user interface (GUI), which provides the following functions in a user-friendly interactive manner: 1) for mining \textit{MLCSs} of input sequences, 2) for further inspecting and analyzing the mined \textit{MLCSs} through the provided intuitive graphs and statistics information, 3) for viewing and downloading the mined results in graph and text forms, and 4) for displaying the sequence samples, whose functions are corresponding to the four radio buttons on the leftmost side of OVT-MLCS GUI in Figure \ref{fig2}.
\vspace{-1em}
\vspace{-1.5em}

\begin{figure*}[!h]
    \centering
    \includegraphics[width=0.99\linewidth]{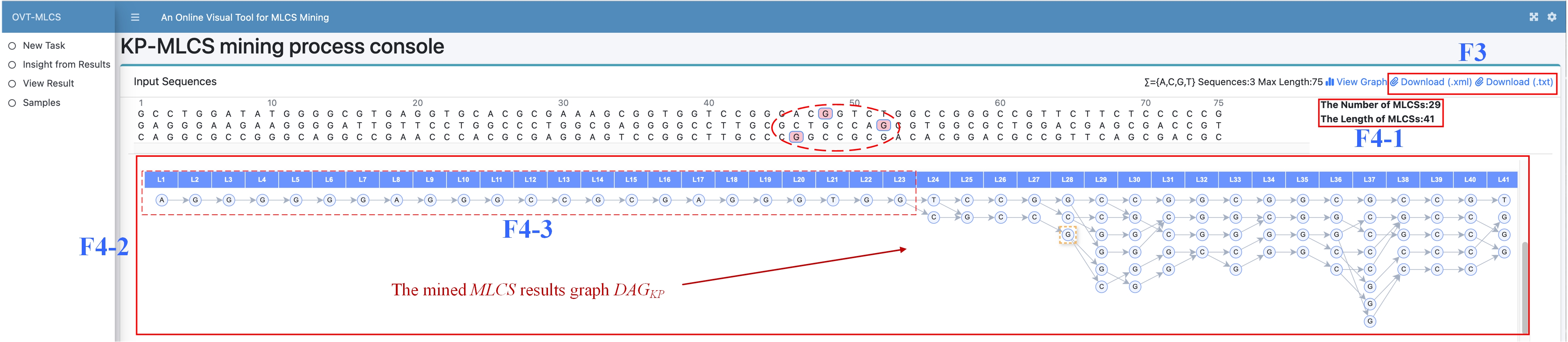}
    \caption{OVT-MLCS mined \textit{MLCS} results $DAG_{KP}$ diagram.}
    \label{fig3}
\end{figure*}

\subsection{Exact/Top-K MLCS Mining}
Contrary to existing \textit{MLCS} algorithms and tools, the proposed OVT-MLCS can quickly and accurately mine all/top-k \textit{MLCSs} for long and big sequences. The details are as follows.

\vspace{0.15em}
\noindent \textbf{Exact MLCS Mining}. This means mining all \textit{MLCSs} from given long or big sequences. To this end, based on our proposed KP-MLCS algorithm \cite{r8}, we introduce the following strategies to overcome the challenging issues faced by existing \textit{MLCS} algorithms/tools (see Section 1, paragraph 3, issue 1), memory explosion or extremely high time complexity: 1) a novel \textit{MLCS} problem-solving graph model, which contains only the nodes and edges that contribute to \textit{MLCS} mining for the given sequences, called $DAG_{KP}$ (shown in the red sub-graph in Figure 6), 2) multi-threaded concurrent \textit{MLCS} mining, and 3) multi-component collaboration of the \textit{Supporting Layer}. Based on the multi-component collaboration of the \textit{Supporting Layer} (see Figure \ref{fig1}), OVT-MLCS can dynamically monitor and manage the memory status during \textit{MLCS} mining. When the memory capacity reaches the upper threshold, OVT-MLCS can control the first several layers of $DAG_{KP}$ subgraphs stored on hard disk DB H2 (called serialization) timely and automatically. When some data in $DAG_{KP}$ are needed to compute or display \textit{MLCS} results, OVT-MLCS read $DAG_{KP}$ in memory layer by layer (called de-serialization).

\vspace{0.15em}
\noindent \textbf{Top-\textit{K} MLCS Mining}. For given input sequences, there are usually multiple \textit{MLCS} solutions, and the number of \textit{MLCS} solutions for long or big sequences can reach tens of thousands or even more \cite{r1,r8}. Therefore, based on practical application requirements, OVT-MLCS can provide top-k \textit{MLCSs}  mining. Specifically, based on the proposed KP-MLCS and its score function \cite{r8} of each node/point in the results graph (denoted by $DAG_{KP}$), it only mines and displays the top-k optimal \textit{MLCSs}, that is, the top-k  \textit{MLCSs} with the least number of discontinuous spaces in the mined \textit{MLCSs}.

\vspace{0.15em}
\textcolor{black}{Note that for the "Exact/Top-K MLCS Mining" of OVT-MLCS, the time/space complexity is $O(dN)+O(E)$ (see \cite{r8} for details), where $N$ and $E$ refer to the total numbers of nodes and edges that need to be processed when constructing $DAG_{KP}$ respectively\textcolor{black}{, while $d$ denotes the number of input sequences.} In general, given 3 long/big sequences as input, "Exact/Top-K MLCS Mining" of OVT-MLCS can be completed in a few to tens of seconds (for long sequences)/minutes (for big sequences). The above two functional buttons and their three corresponding different input modes are shown in the red boxes, F1 and F2, in Figure 2.}

\noindent \textbf{Display of Mined MLCS Results}. To overcome the challenging issue faced by existing \textit{MLCS} algorithms/tools (see Section 1, paragraph 3, issue 2), OVT-MLCS compresses and displays all mined \textit{MLCSs} at once in the form of a $DAG_{KP}$ graph, in which each path represents an \textit{MLCS} solution. Moreover, by introducing Antv-X6 open source graphics engine and SVG (\underline{S}calable \underline{V}ector \underline{G}raphics) technology, OVT-MLCS can display, scale up/down and interact with the mined  \textit{MLCS} results graph $DAG_{KP}$ in web pages (see F4-2 in Figure \ref{fig3}, while F4-1 gives the statistical information of $DAG_{KP}$).  

\begin{figure}[!h]
    \centering
    \includegraphics[width=0.99\linewidth]{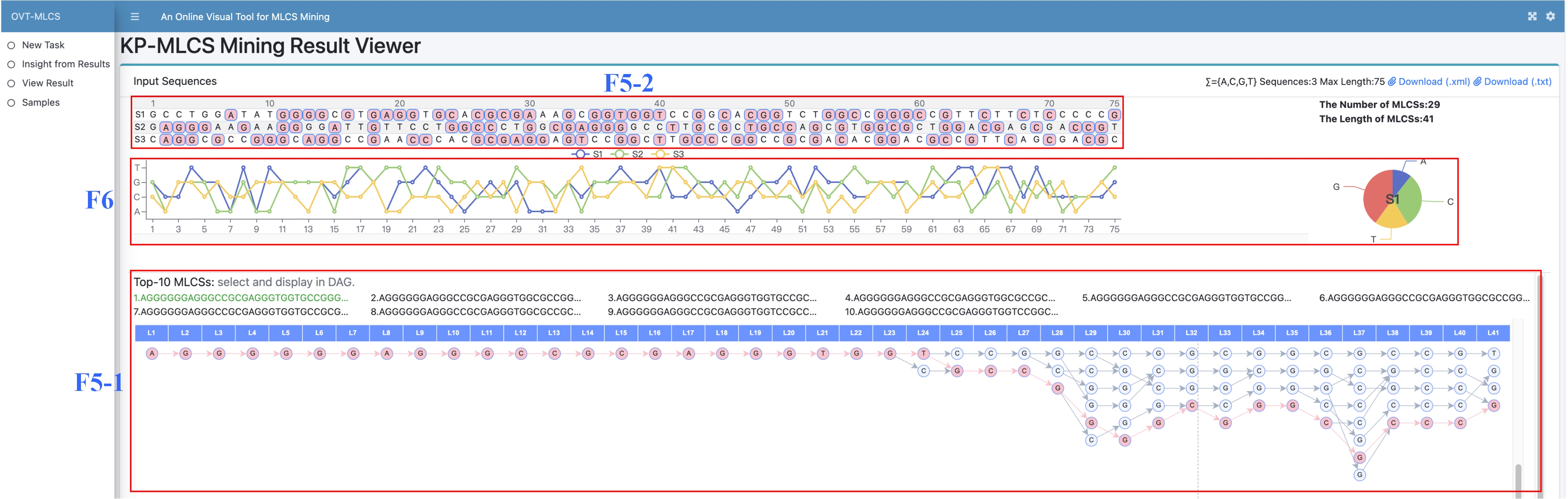}
    \caption{OVT-MLCS Mining Results Insight (a).}
    \label{fig4}
\end{figure}
\vspace{-1.5em}
\begin{figure}[!h]
    \centering
    \includegraphics[width=0.95\linewidth]{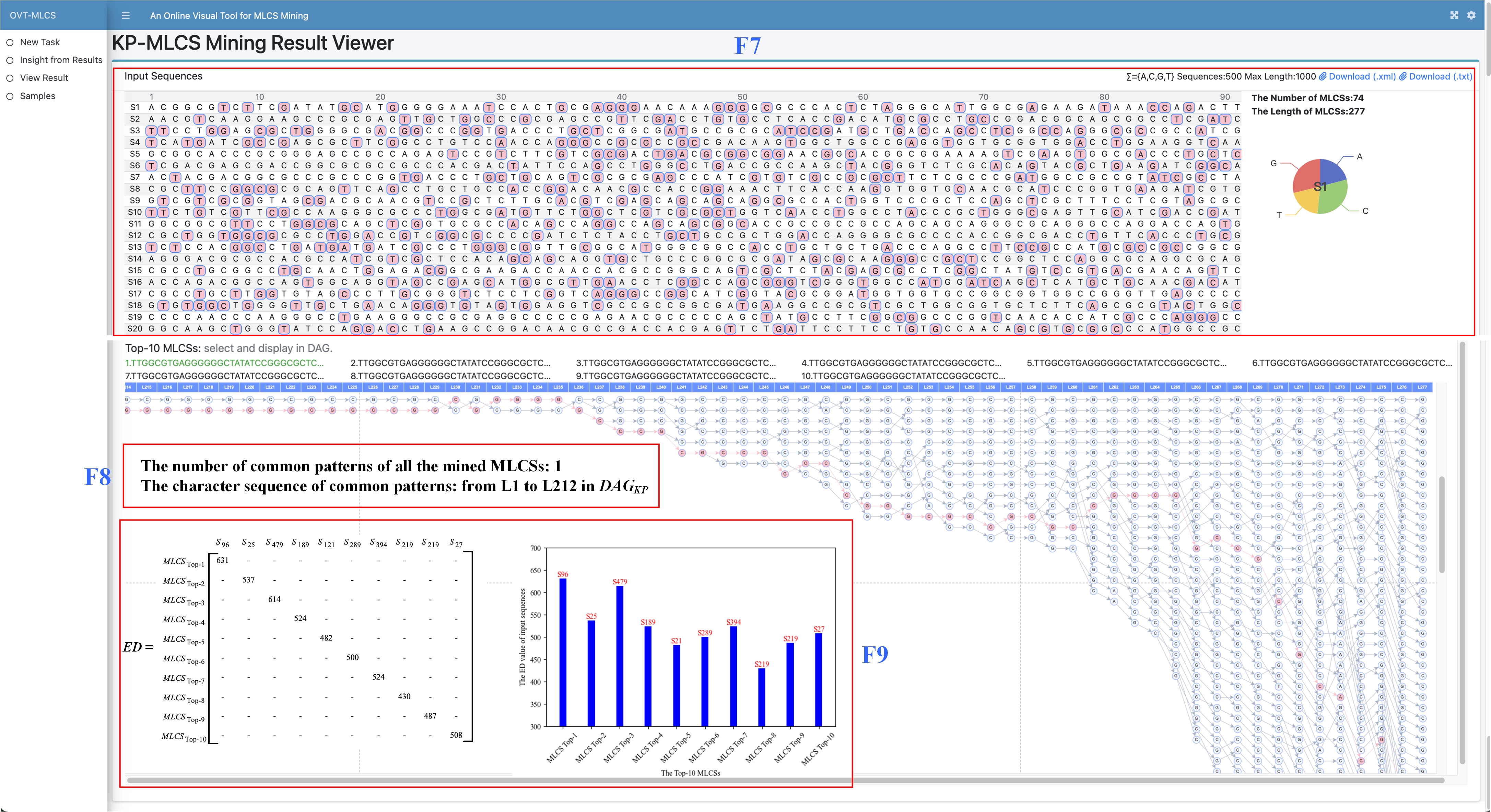}
    \caption{OVT-MLCS Mining Results Insight (b).}
    \label{fig5}
    \vspace{-1em}
\end{figure}

\subsection{Mining Result Insight}

To overcome the challenges of existing \textit{MLCS} algorithms/ tools and to provide mining results insight to users, based on "\textbf{Display of Mined MLCS Results}"(see Section 2.2), OVT-MLCS further presents the following novel ways for displaying and interacting with mining results in a user-friendly manner. That is, 1) OVT-MLCS presents some visual line charts of input sequences and a pie chart showing the proportion of each character in each sequence (see F6 in Figure \ref{fig5}), from which the user can get an intuitive view of the similarities and common patterns of input sequences. 2) OVT-MLCS  automatically displays the top-10 mined \textit{MLCS} results in the forms of graph $DAG_{KP}$ and the corresponding input sequence, \textit{where in the displayed $DAG_{KP}$, the subgraph sections with width 1 clearly reveal the common pattern of the mined \textit{MLCSs} without any additional calculations} (see F4-3 in Figure \ref{fig3}). 3) OVT-MLCS provides an ability for the user to interact with the mined top-10 \textit{MLCSs} accompanied by automatic calculation and simultaneous display of a variety of statistical information (see F8 and F9 in Figure \ref{fig5}) so that the user can further observe/analyze the patterns from both the mined \textit{MLCSs} and input sequences (see F5-1 and F5-2 in Figure \ref{fig4}, or F7 and F8 in Figure \ref{fig5}). It is worth noting that \textit{the two-way online inspection and interaction from the system input to its output are unique and do not exist in existing systems/tools, which will open up new perspectives for practical application analysis}.

\vspace{-0.4em}
\subsection{Mining Result Download}
With the help of the underlying database H2, the components \textit{Mining Memory Manager} (calling the API of \textit{Websocket Excavation Monitor} for monitoring and managing system memory status) and \textit{Mining Results Persistence} (achieving the serialization and de-serialization of mining results), OVT-MLCS provides the download functions to the mined \textit{MLCSs} in two forms, text file (.text) and graph file (.xml) for the user (see F3 red box marked in Figure \ref{fig3}).

\section{Demonstration Overview}

\subsection{Functions and Features of OVT-MLCS}
In this part, we demonstrate and explain the architecture, key functions (i.e., \textit{Exact/Top-K MLCS Mining}, \textit{Display of Mined MLCS Results}, \textit{Mining Result Insight} and \textit{Mining Result Download}), features, and user-interaction modes of OVT-MLCS through several \textit{MLCS} use cases of mining long or big sequences.

\subsection{Practical Use Cases of OVT-MLCS}
Here we demonstrate two \textit{MLCS} mining use cases of OVT-MLCS for big sequences \footnote{The detailed information of the big sequences using the following two practical use cases are available at https://github.com/OVT-MLCS/OVT-MLCS-Tool} in biomedical scenarios. It should be noted that \textit{there is no existing \textit{MLCS} algorithm/tool to carry out \textit{MLCS} mining for big sequences yet}.

\vspace{0.25em}
\noindent \textbf{Practical Use Case 1}: A biomedical user collected many publicly available complete genome sequences of the COVID-19 virus from different countries, and related flu-causing coronaviruses/viruses. These are big sequences with a length of about 30,000 in alphabet $\Sigma=\{A, C, G, T\}$. For research and development of relevant vaccines and targeted drugs, the user wants to obtain 1) the evolutionary relationships of COVID-19 viruses and 2) the similarities between COVID-19 viruses and various flu-causing coronaviruses/viruses.

We will demonstrate that by adopting functions \textit{Exact/Top-K Mining} and \textit{Mining Result Insight} in OVT-MLCS, the user can fulfill his/her needs {\textcolor{black}{in 1.5 hours}}.

\vspace{0.25em}
\noindent \textbf{Practical Use Case 2}: A biomedical user selected 11 publicly available complete genome sequences (length $\ge 10000$) from liver cancer patients based on the latest sequencing technology. To support accurate early cancer gene testing, personalized treatment guidance, and cancer prevention, the user wants to 1) discover new liver cancer mutation target locations and 2) observe/analyze the commonality (i.e., common pattern) and individuality of the mutation target locations of these liver cancer patients.

We will demonstrate that by using \textit{Top-K Mining} of OVT-MLCS, inspecting common patterns directly in the mined results $DAG_{KP}$, and using two-way online inspection and interaction between input sequences and the mined results $DAG_{KP}$ (see Section 2.3 for details), the user can obtain his/her expected results {\textcolor{black}{in 25  minutes}}.


\vspace{-0.5em}
\section{Related Algorithms and Systems}

\begin{figure}[!t]
    \centering
    \includegraphics[width=0.75\linewidth,height=0.13\paperheight]{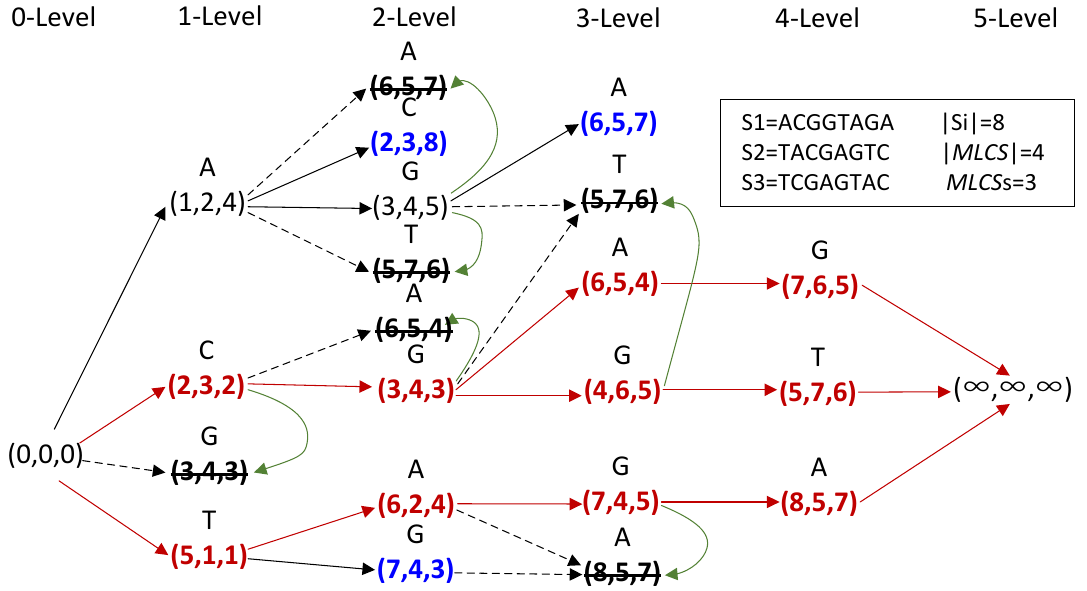}
    \vspace{-2ex}\caption{The constructed \textit{MLCS-DAG} of $S_1$,  $S_2$ and $S_3$ over $\Sigma=\{A, C, G, T\}$ by a DOP-based algorithm, where both the blue points and points with strikethroughs denote non-critical points, which together with their corresponding arrows, should be deleted as they don’t contribute to \textit{MLCS} mining. All the \textit{MLCS}s (linked by red arrows, $MLCS_1=CGAG$, $MLCS_2=CGGT$ and $MLCS_3=TAGA$ of length 4, i.e., $|MLCS|=4$) can be obtained by tracing back from the sink point $(\infty, \infty, \infty)$ to the source point $(0, 0, 0)$ on the \textit{MLCS-DAG}.}
    \vspace{-1em}
    \label{fig6}
\end{figure}

\noindent \textbf{Related Algorithms}. Existing \textit{MLCS} algorithms can be divided into two main categories: the classic \textit{dynamic programming (DYP) based} algorithms, which have a high space-time complexity, and the current leading \textit{dominant-point (DOP) based} algorithms, which have a lower space-time complexity than that of the DYP based algorithms. Figure \ref{fig6}  depicts the constructed \textit{MLCS-DAG} and the mined \textit{MLCSs} of sequences $S_1$,  $S_2$ and $S_3$ over $\Sigma=\{A, C, G, T\}$ using a DOP-based algorithm. Reference \cite{r8} reveals that the existing leading DOP-based \textit{MLCS} algorithms have the following drawbacks: 1) they need intensive calculations and produce useless nodes in the constructed MLCS-DAG graph, which makes it impossible to carry out \textit{MLCSs} mining for long or big sequences, and 2) traversing through the \textit{MLCS-DAG} to find all the mining results, \textit{MLCSs}, one by one is not only time-consuming but also hinders the inspection and discovery of patterns in the mining results. In contrast to all existing \textit{MLCS} algorithms, we propose a new \textit{MLCS} problem-solving graph model $DAG_{KP}$, which contains only key points (i.e., the red sub-graph in Figure \ref{fig6}), and a novel parallel \textit{MLCS} algorithm called KP-MLCS \cite{r8}, which can mine and compress all \textit{MLCSs} of long or big sequences effectively and efficiently. It is worth noting that to meet the functional requirements of our OVT-MLCS, based on our KP-MLCS, we propose some novel methods/technologies, such as online Top-K \textit{MLCSs} exact mining, online {\textcolor{black}{visualization, compression and serialization of the \textit{MLCSs} graph $DAG_{KP}$, intuitive common pattern representation, etc., which contribute to making our KP-MLCS method more powerful and \textcolor{black}{practical}.}  

\vspace{0.5em}
\noindent \textbf{Related Tools}. There are several well-known sequence alignment tools, e.g., BLAST \cite{r10}, Clustal Omega \cite{r11}, etc. BLAST (\underline{B}asic \underline{L}ocal \underline{A}lignment \underline{S}earch \underline{T}ool) is a commonly used tool in bioinformatics, which can compare the input nucleic acid or protein sequence with a known sequence in the database to obtain information such as sequence similarity to determine the origin or evolutionary relationship of the sequences. Clustal Omega is a software for the alignment of multiple sequences. Our OVT-MLCS is different from the above tools in the following aspects: 1) it is the first \textit{MLCS} mining tool focused on long or big character sequences with a scale of 3 to 5000, that is, not limited to biological sequences, and 2) it has a comprehensive and user-friendly online visual tool for long or large sequence alignment, similarity comparison, and in-depth analysis of common patterns in input sequences or their \textit{MLCSs}.

\vspace{-0.5em}
\begin{acks}
This work of Zhi Wang, Yanni Li and Hui Li is partially supported by the National Natural Scientific Foundation of China (Grant No. 62176202 and 62272369).
\end{acks}

\bibliographystyle{ACM-Reference-Format}
\bibliography{refs}

\end{document}